\documentclass[english,10pt,letterpaper]{article}
\usepackage[T1]{fontenc}
\usepackage[left=35mm, right=35mm, top=30mm, bottom=30mm]{geometry}
\usepackage{graphicx}
\usepackage{authblk}
\usepackage{amssymb}
\usepackage{amsmath}
\usepackage{babel}
\usepackage{indentfirst}
\usepackage{setspace}
\usepackage{lettrine}
\usepackage{fancyhdr}
\usepackage{stackengine}
\usepackage[labelfont=bf]{caption}
\providecommand{\indexterms}[1]{\small{\textbf{\textit{Index Terms---}}}#1}
\providecommand{\abst}[1]{\textbf{\textit{Abstract---}}#1}

\title{\Huge{\textbf{Analysis of Stability in Multistage Feedforward Operational Transconductance Amplifiers using Successive One-Pole Approximation}}}
\author{Taeju Lee}
\affil{\small{Department of Electrical Engineering, Columbia University, New York, NY 10027, USA\\
	E-mail: taeju.lee@columbia.edu}}
\date{}

\begin{document}
\maketitle
\pagestyle{fancy}
\fancyhf{}
\lhead{\small{LEE: ANALYSIS OF STABILITY IN MULTISTAGE FEEDFORWARD OTAs USING SOPA}}
\cfoot{\thepage}

\noindent
\small{\abst{\textbf{A high-gain wideband operational transconductance amplifier (OTA) is an essential block for applications requiring high data rates. As the CMOS technology node has scaled down, the minimum length of transistors has shrunk and the intrinsic gain has been reduced. Therefore, design techniques that achieve a high-gain wideband operation are essential in analog amplifier design using advanced CMOS technology nodes. In analog amplifier design, the stability issue can arise as the gain and the bandwidth are improved, degrading the overall performance of analog circuits. Accordingly, appropriate frequency compensation is required to ensure the reliable and precise operation of an OTA. This paper presents the stability analysis of OTAs that employ the feedforward path and the multistage amplifier to achieve high-gain wideband operation. To analyze the stability of multistage feedforward OTAs and provide an intuitive design method, the successive one-pole approximation (SOPA) is applied to each substage of the multistage feedforward OTA. By employing SOPA, the stability analysis is systematically carried out from the two-stage feedforward OTA to the four-stage feedforward OTA in this work. Also, the design limitation of increasing the number of stages is discussed based on SOPA.}}}
\singlespacing
\noindent
\small{\indexterms{\textbf{Analog front-end, amplifier, CMOS, compensation technique, feedforward, frequency response, high gain, high speed, multistage, operational transconductance amplifier (OTA), stability, successive one-pole approximation (SOPA), wideband.}}}

\singlespacing
\singlespacing
\singlespacing
\singlespacing

\section{\Large{I}\large{NTRODUCTION}}
\lettrine[findent=2pt, nindent=0pt]{\textbf{S}}{\textbf{emiconductor}} technology has continued to develop, which has significantly advanced electronic devices in terms of power, speed, and size. As CMOS technology has advanced rapidly, computing engines and communication devices with high data rates become available. In these applications that require high data rates, a high-gain wideband operational transconductance amplifier (OTA) is essential for various analog blocks. As the technology node shrinks, digital circuits can be developed to occupy less area, consume lower power, and enable high-speed operation. However, when developing analog circuits such as OTAs using advanced CMOS technologies, they must be carefully designed while considering the performance such as stability and gain.

\singlespacing
As CMOS technology has advanced, the minimum length of transistors has decreased, which has improved the speed and power efficiency of digital circuits such as flip-flops and logic gates. However, an intrinsic gain of a single transistor is degraded due to the minimized gate length, causing gain errors in negative feedback systems. To overcome this gain degradation, the OTA can be developed by cascading multiple gain stages \cite{r1}--\cite{r5}. Especially, a two-stage OTA has been widely employed in bias circuits, analog front-ends, analog-to-digital converters, and wireless receivers for applications such as communications \cite{r6}--\cite{r9} and sensor interfaces \cite{r10}--\cite{r13}.

As the number of cascaded gain stages increases, additional poles are created, causing phase shift and degrading stability in feedback systems. Therefore, in an OTA used in a feedback system, the technique called Miller compensation has been widely used to compensate for stability by controlling phase shift \cite{r4}, \cite{r5}. Among the diverse compensation techniques to ensure stability in multiple gain stages, including the Miller compensation technique, the feedforward structure can be employed as a frequency compensation solution in a multistage OTA, which improves stability by canceling out poles \cite{r1}. This paper provides the analysis of stability in multistage feedforward OTAs which achieve high-gain wideband operation. In particular, the successive one-pole approximation (SOPA) is developed in this work to conduct intuitive stability analysis in multistage feedforward OTAs. By utilizing SOPA, systematic stability analysis is possible in feedforward OTAs with increasing gain stages. SOPA means that the one-pole approximation is successively applied to each substage of the multistage feedforward OTA, which is introduced in the remaining work of this paper. In this paper, by employing SOPA, the stability analysis is carried out from the two-stage feedforward OTA to the four-stage feedforward OTA. Additionally, the design limitation of increasing the number of feedforward stages is described based on SOPA after discussing the four-stage feedforward OTA.

This paper is organized as follows. Section II reviews Miller compensation which is a widely used method to secure stability for feedback systems. Section III provides the stability analysis by employing SOPA from the two-stage feedforward OTA to the four-stage feedforward OTA. Finally, the conclusion is drawn in Section IV. All analysis results are verified using a 65-nm CMOS technology.

\singlespacing
\singlespacing

\section{\Large{G}\large{ENERAL} \Large{C}\large{ONSIDERATION}}
\begin{figure}[h!]
	\centering
	\includegraphics[scale=0.5]{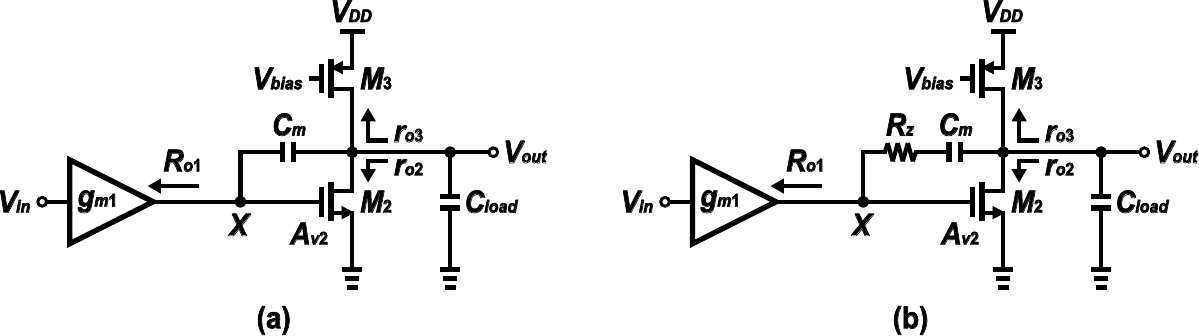}
	\caption{(a) Miller compensation using $C\textsubscript{\textit{m}}$. (b) Addition of $R\textsubscript{\textit{z}}$ to remove the RHP zero.}
	\label{Fig. 1}
\end{figure}

Two-stage OTAs are widely used to provide a sufficient gain, but an additional pole contributes to phase shift, ultimately degrading the phase margin (PM). To mitigate the PM degradation, Miller compensation has been widely employed in numerous OTA designs. As shown in Fig. \ref{Fig. 1}(a), the Miller capacitor $C\textsubscript{\textit{m}}$ is placed between the input and output nodes of the second stage $M\textsubscript{2}$, splitting poles and ultimately improving the PM. At the interstage node $X$, the capacitance looking into the gate terminal of $M\textsubscript{2}$ increases to $(1+A\textsubscript{\textit{v}2})C\textsubscript{\textit{m}}$, shifting the dominant pole toward the origin in low frequencies \cite{r4}. When looking into the output node of $M\textsubscript{2}$ at very high frequencies, the resistance decreases from $r\textsubscript{\textit{o}2}||r\textsubscript{\textit{o}3}$ to approximately $R\textsubscript{\textit{o}1}||(1/g\textsubscript{\textit{m}2})||r\textsubscript{\textit{o}2}||r\textsubscript{\textit{o}3} \approx 1/g\textsubscript{\textit{m}2}$ since $C\textsubscript{\textit{m}}$ becomes a low-impedance path at high frequencies \cite{r4}. Accordingly, when $C\textsubscript{\textit{m}}$ is employed, the pole at node $X$ moves toward the origin, the pole at the output node moves away from the origin, and two poles are split \cite{r4}. $R\textsubscript{\textit{o}1}$ means the output resistance of the first stage. $r\textsubscript{\textit{o}2}$ and $r\textsubscript{\textit{o}3}$ are the output resistances of $M\textsubscript{2}$ and $M\textsubscript{3}$, respectively.

However, employing $C\textsubscript{\textit{m}}$ also provides the feedforward path from node $X$ to the output node, contributing to a negative phase shift. In other words, as $C\textsubscript{\textit{m}}$ creates the feedforward path, the gain becomes zero at $\omega\textsubscript{\textit{z}}$ called a right-half-plane (RHP) zero. In Fig. \ref{Fig. 1}(a), the RHP zero is obtained by solving the condition of $V\textsubscript{\textit{X}}C\textsubscript{\textit{m}}\omega\textsubscript{\textit{z}}=V\textsubscript{\textit{X}}g\textsubscript{\textit{m}2}$ which means that the small-signal output becomes zero at $\omega\textsubscript{\textit{z}}$ \cite{r4}. Considering the gate-drain parasitic capacitance of $M\textsubscript{2}$, $C\textsubscript{\textit{gd}2}$, $\omega\textsubscript{\textit{z}}$ can be rewritten as $g\textsubscript{\textit{m}2}/(C\textsubscript{\textit{m}}+C\textsubscript{\textit{gd}2})$. The RHP zero contributes to a negative phase shift as with a pole, resulting in the PM degradation. As a solution to remove the RHP zero, the compensation resistor $R\textsubscript{\textit{z}}$ can be series-connected with $C\textsubscript{\textit{m}}$ as shown in Fig. \ref{Fig. 1}(b). After adding $R\textsubscript{\textit{z}}$, $\omega\textsubscript{\textit{z}}$ is obtained as $1/[C\textsubscript{\textit{m}}(g\textsubscript{\textit{m}2}^{-1}-R\textsubscript{\textit{z}})]$ by solving $V\textsubscript{\textit{X}}g\textsubscript{\textit{m}2}=V\textsubscript{\textit{X}}[R\textsubscript{\textit{z}}+(C\textsubscript{\textit{m}}\omega\textsubscript{\textit{z}})^{-1}]^{-1}$ \cite{r4}. Therefore, the RHP zero can move to the left-half plane by satisfying $R\textsubscript{\textit{z}} > g\textsubscript{\textit{m}2}^{-1}$. The left-half-plane (LHP) zero contributes to a positive phase shift, which can cancel out a pole through appropriate design techniques and eventually improve the phase margin and stability.

As a solution to secure the PM and stability, $C\textsubscript{\textit{m}}$ and $R\textsubscript{\textit{z}}$ can be employed for performing Miller compensation \cite{r4}, and stability can be also secured by removing poles by adding active feedforward paths \cite{r1}. The following stability analysis covers the compensation technique with LHP zeros generated by feedforward paths implemented using active devices. To conduct a systematical stability analysis in multistage feedforward OTAs, SOPA is applied to each design of multistage feedforward OTAs. For simplicity, the stability analysis is first conducted on the two-stage feedforward structure using the one-pole approximation, and SOPA is explained based on the results of the two-stage analysis. Then, the stability analysis of three- and four-stage structures is carried out using SOPA. The design limitation of increasing the number of feedforward stages is also discussed using SOPA at the end of this work.

\singlespacing
\singlespacing

\section{Stability Analysis using SOPA}
\subsection{Analysis of Two-Stage Feedforward Structure}
\begin{figure}[h!]
	\centering
	\includegraphics[scale=0.5]{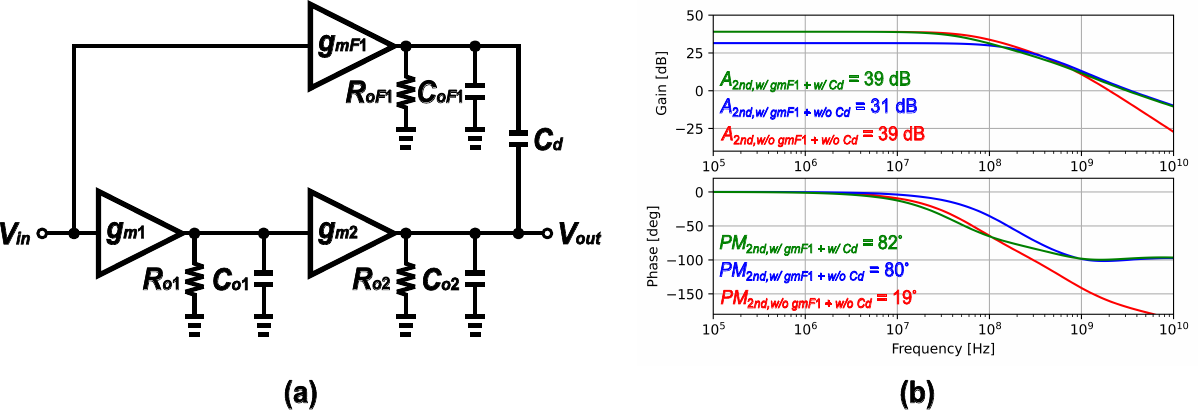}
	\caption{(a) Two-stage feedforward OTA and (b) its Bode plots.}
	\label{Fig. 2}
\end{figure}

For simplicity of multistage stability analysis, a two-stage feedforward OTA is first analyzed using the diagram shown in Fig. \ref{Fig. 2}(a). When neglecting the decoupling capacitor $C\textsubscript{\textit{d}}$, the gains of the main path are expressed as $A\textsubscript{\textit{v}1}=g\textsubscript{\textit{m}1}R\textsubscript{\textit{o}1}$ and $A\textsubscript{\textit{v}2}=g\textsubscript{\textit{m}2}(R\textsubscript{\textit{o}2}||R\textsubscript{\textit{oF}1})$. The gain of the feedforward path is $A\textsubscript{\textit{vF}1}=g\textsubscript{\textit{mF}1}(R\textsubscript{\textit{o}2}||R\textsubscript{\textit{oF}1})$. Then, the overall transfer function of the two-stage feedforward structure is given by

\begin{align}
	\frac{V\textsubscript{\textit{out}}}{V\textsubscript{\textit{in}}}(s)\bigg|\textsubscript{2-\textit{stage}}&=\frac{A\textsubscript{\textit{v}1}A\textsubscript{\textit{v}2}}{\Big(1+\frac{s}{\omega\textsubscript{\textit{p}1}}\Big)\Big(1+\frac{s}{\omega\textsubscript{\textit{p}2}}\Big)}+\frac{A\textsubscript{\textit{vF}1}}{\Big(1+\frac{s}{\omega\textsubscript{\textit{p}2}}\Big)}\\
	\nonumber
	\\
	\nonumber
	&=\big(A\textsubscript{\textit{v}1}A\textsubscript{\textit{v}2}+A\textsubscript{\textit{vF}1}\big)\\
	&\hspace{10pt}\times\frac{\bigg[1+\frac{A\textsubscript{\textit{vF}1}s}{\big(A\textsubscript{\textit{v}1}A\textsubscript{\textit{v}2}+A\textsubscript{\textit{vF}1}\big)\omega\textsubscript{\textit{p}1}}\bigg]}{\Big(1+\frac{s}{\omega\textsubscript{\textit{p}1}}\Big)\Big(1+\frac{s}{\omega\textsubscript{\textit{p}2}}\Big)}
\end{align}

The multistage feedforward OTAs used in this work are designed based on \cite{r1}. Eqs. (1) and (2) are obtained in the same way as \cite{r1}. From Eq. (2), the voltage gain at low frequencies can be obtained as $g\textsubscript{\textit{m}1}R\textsubscript{\textit{o}1}g\textsubscript{\textit{m}2}(R\textsubscript{\textit{o}2}||R\textsubscript{\textit{oF}1})+g\textsubscript{\textit{mF}1}(R\textsubscript{\textit{o}2}||R\textsubscript{\textit{oF}1})$ and approximated as $g\textsubscript{\textit{m}1}R\textsubscript{\textit{o}1}g\textsubscript{\textit{m}2}(R\textsubscript{\textit{o}2}||R\textsubscript{\textit{oF}1})$. In Fig. \ref{Fig. 2}(a), if $C\textsubscript{\textit{d}}$ is neglected, the poles at the first and second output stages are obtained as $\omega\textsubscript{\textit{p}1}=1/(R\textsubscript{\textit{o}1}C\textsubscript{\textit{o}1})$ and $\omega\textsubscript{\textit{p}2}=1/(R\textsubscript{\textit{o}2}||R\textsubscript{\textit{oF}1})(C\textsubscript{\textit{o}2}+C\textsubscript{\textit{oF}1})$. From Eq. (2), the LHP zero generated by the feedforward path of $g\textsubscript{\textit{mF}1}$ is expressed as $\omega\textsubscript{\textit{z}1}=\omega\textsubscript{\textit{p}1}[(A\textsubscript{\textit{v}1}A\textsubscript{\textit{v}2}+A\textsubscript{\textit{vF}1})/A\textsubscript{\textit{vF}1}]=\omega\textsubscript{\textit{p}1}[(g\textsubscript{\textit{m}1}R\textsubscript{\textit{o}1}g\textsubscript{\textit{m}2}/g\textsubscript{\textit{mF}1})+1]$. If $g\textsubscript{\textit{mF}1} \gg g\textsubscript{\textit{m}1},g\textsubscript{\textit{m}2}$, $\omega\textsubscript{\textit{z}1}$ can be approximated as $\omega\textsubscript{\textit{p}1}$. Therefore, the first pole $\omega\textsubscript{\textit{p}1}$ can be canceled out by choosing $g\textsubscript{\textit{mF}1} \gg g\textsubscript{\textit{m}1},g\textsubscript{\textit{m}2}$. This ultimately improves the stability by converting the two-pole system into the one-pole system, which is called the one-pole approximation. Then, SOPA is an extended version of the one-pole approximation.

Although the PM is improved by employing a feedforward path, the voltage gain is degraded by $R\textsubscript{\textit{oF}1}$. Therefore, to isolate the output resistance of the feedforward path from the main path, the decoupling capacitor $C\textsubscript{\textit{d}}$ can be placed between the feedforward path output and the main path output \cite{r2}. Compared to the design that places $C\textsubscript{\textit{d}}$ at all inputs and outputs of feedforward paths \cite{r2}, this work places $C\textsubscript{\textit{d}}$ only at the output of the outer feedforward path to save the design area. By placing $C\textsubscript{\textit{d}}$ as shown in Fig. \ref{Fig. 2}(a), the output impedance of the feedforward path becomes high at low frequencies and the effective output resistance of the OTA is dominated by $R\textsubscript{\textit{o}2}$. Therefore, the voltage gain changes from $g\textsubscript{\textit{m}1}R\textsubscript{\textit{o}1}g\textsubscript{\textit{m}2}(R\textsubscript{\textit{o}2}||R\textsubscript{\textit{oF}1})$ to $g\textsubscript{\textit{m}1}R\textsubscript{\textit{o}1}g\textsubscript{\textit{m}2}R\textsubscript{\textit{o}2}$. Also, the dominant pole $\omega\textsubscript{\textit{p}2}$ approximately changes from $1/(R\textsubscript{\textit{o}2}||R\textsubscript{\textit{oF}1})(C\textsubscript{\textit{o}2}+C\textsubscript{\textit{oF}1})$ to $1/(R\textsubscript{\textit{o}2}C\textsubscript{\textit{o}2})$ as the output impedance of the feedforward path is isolated from the main path. As the frequency increases, $C\textsubscript{\textit{d}}$ operates almost as a short circuit. Accordingly, $\omega\textsubscript{\textit{p}1}$ and $\omega\textsubscript{\textit{z}1}$ located at high frequencies are almost the same as in the case without $C\textsubscript{\textit{d}}$.

\begin{figure}[h!]
	\centering
	\includegraphics[scale=0.5]{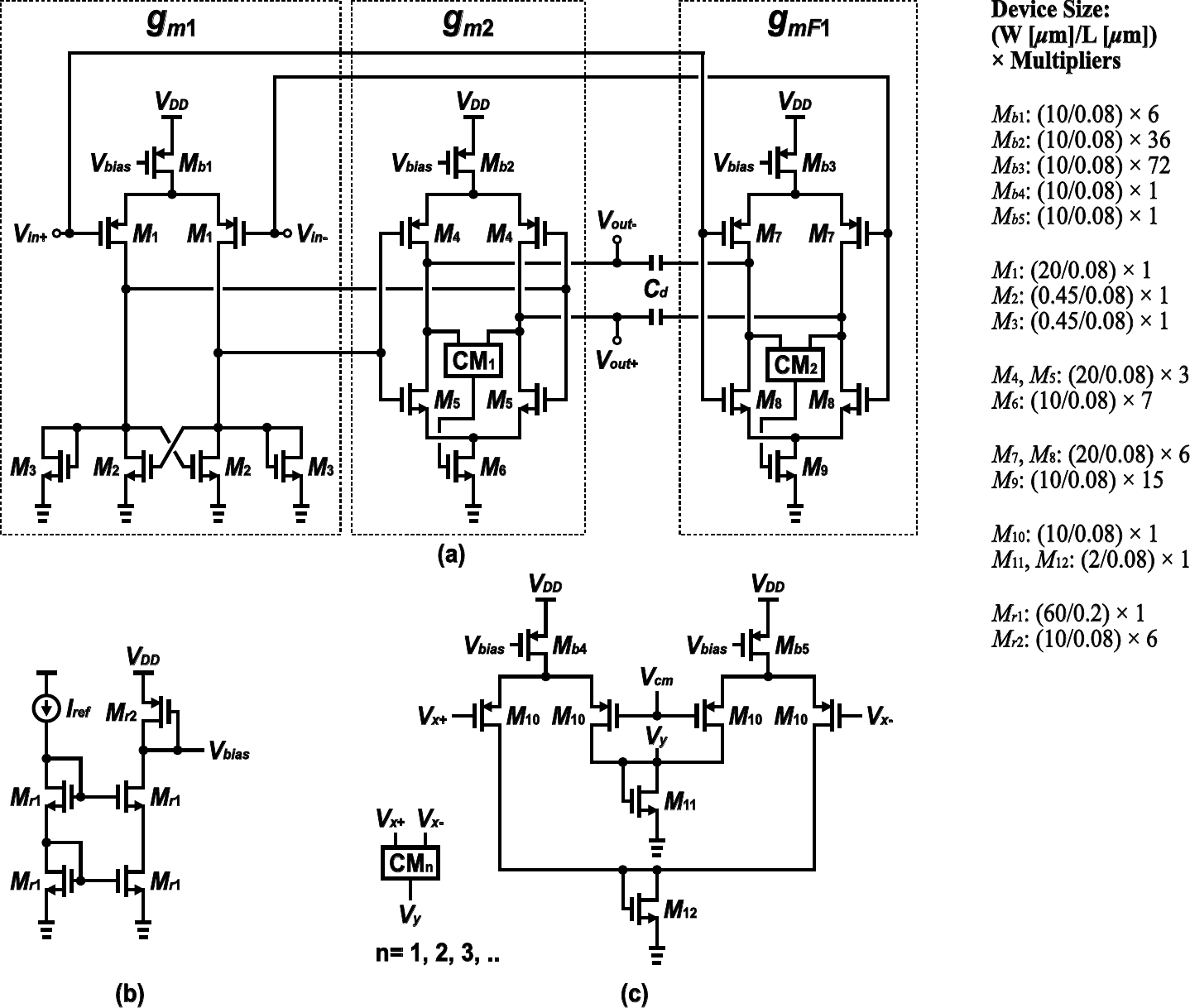}
	\caption{(a) Fully differential two-stage feedforward OTA, (b) its bias block, and (c) the common-mode feedback used in the output stage.}
	\label{Fig. 3}
\end{figure}

Fig. \ref{Fig. 2}(b) shows the simulation results of gain and PM based on the two-stage feedforward OTA shown in Fig. \ref{Fig. 2}(a): using only the two-stage OTA without the feedforward path significantly degrades the PM while achieving a high gain (red line), adding the feedforward path to the main path improves the PM but the gain is reduced due to the output resistance of the feedforward path (blue line), and combining the feedforward path with $C\textsubscript{\textit{d}}$ compensates for the gain while ensuring sufficient PM (green line). As shown in the phase shift of Fig. \ref{Fig. 2}(b), the two-pole system (red line) is approximated as the one-pole system (blue and green lines) as the feedforward path is added.

Fig. \ref{Fig. 2}(b) is obtained using the schematic shown in Fig. \ref{Fig. 3} by setting $C\textsubscript{\textit{d}}$ = 2 pF, $V\textsubscript{\textit{DD}}$ = 1.2 V, $V\textsubscript{\textit{cm}}$ = 0.6 V, and $I\textsubscript{\textit{ref}}$ = 400 $\mu$A. The output load impedance is modeled as 10 M$\Omega||$2 pF on $V\textsubscript{\textit{out}--}$ and $V\textsubscript{\textit{out}+}$. For simplicity of the analysis throughout this work, the equation derivation is conducted using the single-ended schematic (e.g., Fig. \ref{Fig. 2}(a)), and the simulation results are obtained using the fully-differential schematic (e.g., Fig. \ref{Fig. 3}).

Fig. \ref{Fig. 3} shows the overall schematics of the fully differential two-stage feedforward OTA. The input stage $g\textsubscript{\textit{m}1}$ is implemented by employing the cross-coupled structure $M\textsubscript{2}$ and the diode-connected load $M\textsubscript{3}$, which improves the voltage gain by maximizing the output resistance of the first stage \cite{r3}. The output stages of $g\textsubscript{\textit{m}2}$ and $g\textsubscript{\textit{mF}1}$ are implemented as the current reuse (or inverter) structure. The common-mode voltage of each output stage is set using the common-mode feedbacks $CM\textsubscript{1}$ and $CM\textsubscript{2}$. $g\textsubscript{\textit{m}1}$ consumes 273 $\mu$A. $g\textsubscript{\textit{m}2}$ and $g\textsubscript{\textit{mF}1}$ consume 1.48 mA and 3 mA, respectively. Each $CM\textsubscript{\textit{n}=1,2}$ consumes 104 $\mu$A. Throughout this paper, the multistage feedforward OTAs are implemented by employing the bias block and the common-mode feedback, as shown in Figs. \ref{Fig. 3}(b) and (c). Note that the values of transconductances such as $g\textsubscript{\textit{m}1}$, $g\textsubscript{\textit{m}2}$, and $g\textsubscript{\textit{mF}1}$ are summarized in the table at the end of this work.

\begin{figure}[h!]
	\centering
	\includegraphics[scale=0.5]{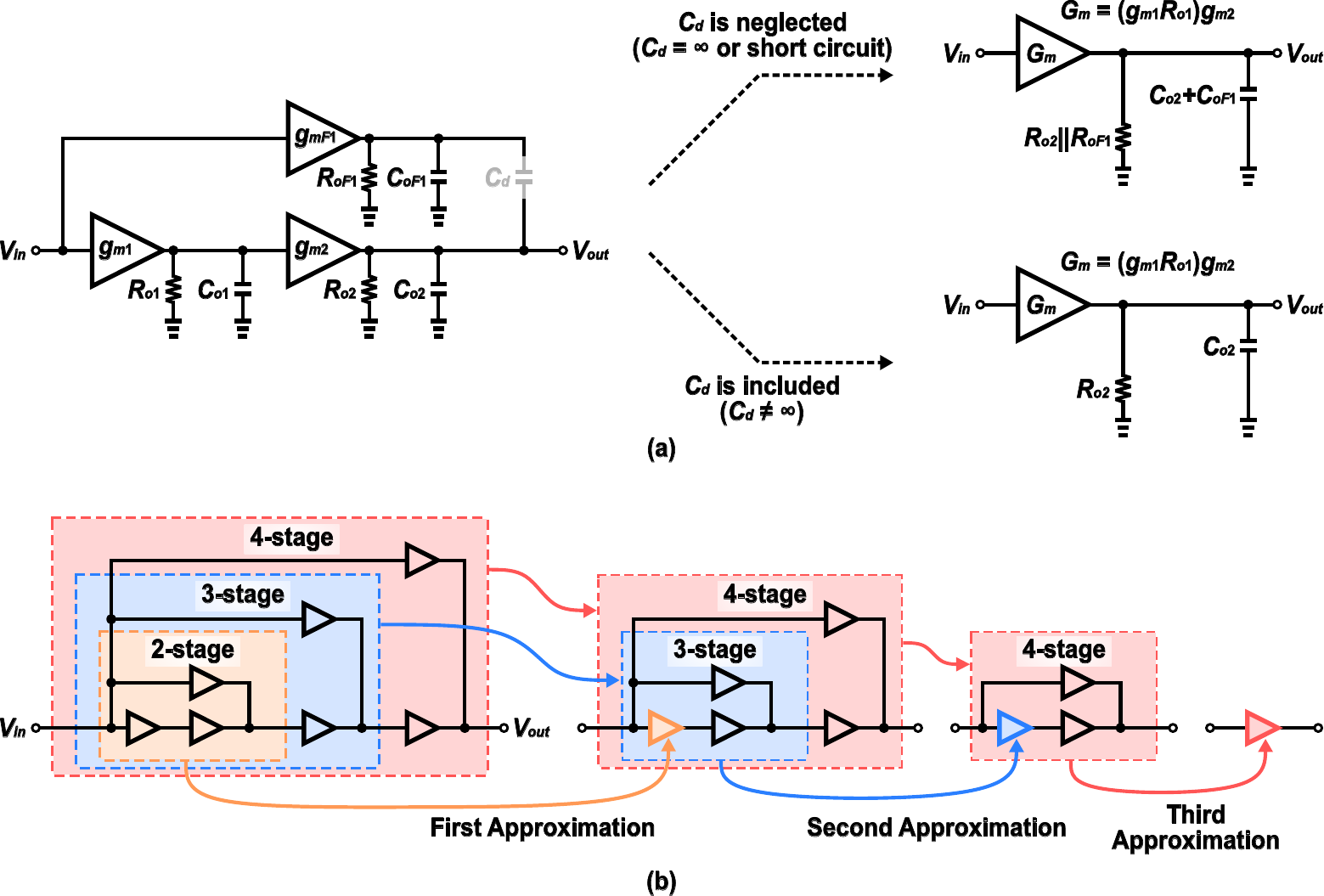}
	\caption{(a) Approximation from the two-pole system to the one-pole system without (or with) $C\textsubscript{\textit{d}}$. (b) Successive one-pole approximation (SOPA).}
	\label{Fig. 4}
\end{figure}

As discussed above using the schematic of Fig. \ref{Fig. 2}(a), the two-pole system can be converted into the one-pole system by choosing $g\textsubscript{\textit{mF}1}$ sufficiently large as shown in Fig. \ref{Fig. 4}(a). To carry out the systematical stability analysis of multistage feedforward OTAs, the one-pole approximation shown in Fig. \ref{Fig. 4}(a) is successively applied to the multistage feedforward structures, defined as the successive one-pole approximation (SOPA) in this work. Fig. \ref{Fig. 4}(a) describes the one-pole approximation depending on $C\textsubscript{\textit{d}}$, which is possible by canceling out the pole using the LHP zero through the feedforward path as \cite{r1}--\cite{r3}, \cite{r7}.

Fig. \ref{Fig. 4}(b) describes the concept of SOPA that approximates the four-stage feedforward OTA into the one-pole system. SOPA is conducted by assuming that each feedforward path of each substage is designed to have a sufficiently large transconductance than that of the main path in the substage. Based on SOPA, the stability analysis of the three- and four-stage feedforward structures is carried out in the remaining work. Note that the substages in Fig. \ref{Fig. 4}(b) are distinguished using different color boxes.

\subsection{Analysis of Three-Stage Feedforward Structure}
\begin{figure}[h!]
	\centering
	\includegraphics[scale=0.5]{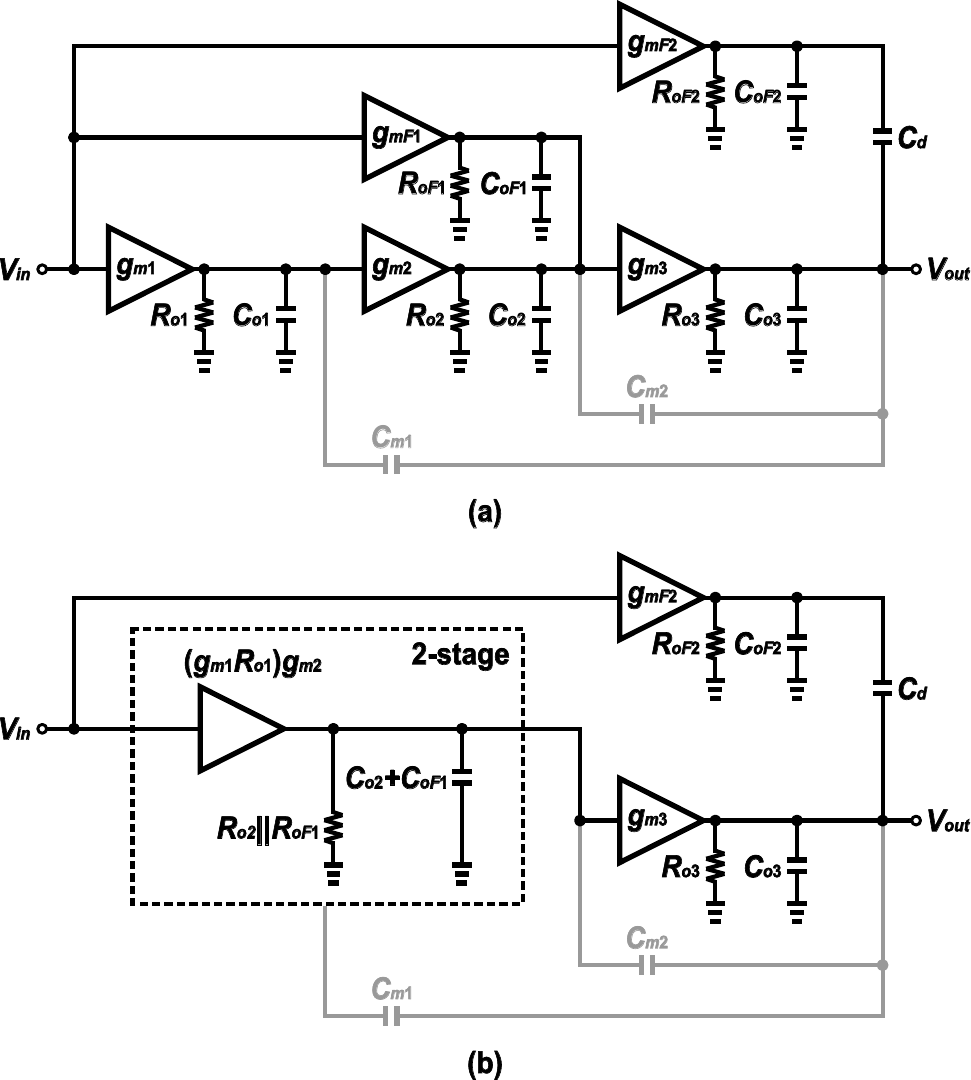}
	\caption{(a) Three-stage feedforward OTA and (b) its approximated schematic.}
	\label{Fig. 5}
\end{figure}

Extending Eqs. (1) and (2) to a three-stage structure, the OTA produces three poles and two zeros in Fig. \ref{Fig. 5}(a). However, obtaining poles and zeros becomes complex as the number of stages extends. Therefore, the three-stage feedforward OTA is approximated by replacing the internal two-stage feedforward structure with the one-pole system as shown in Fig. \ref{Fig. 5}(b). As discussed in the two-stage feedforward OTA shown in Fig. \ref{Fig. 2}(a), $\omega\textsubscript{\textit{p}1}$ can be canceled out by $\omega\textsubscript{\textit{z}1}$ through a feedforward path. Accordingly, a two-stage feedforward OTA can be approximated as a one-pole system with a pole of $1/(R\textsubscript{\textit{o}2}||R\textsubscript{\textit{oF}1})(C\textsubscript{\textit{o}2}+C\textsubscript{\textit{oF}1})$ and a gain of $g\textsubscript{\textit{m}1}R\textsubscript{\textit{o}1}g\textsubscript{\textit{m}2}(R\textsubscript{\textit{o}2}||R\textsubscript{\textit{oF}1})$ as shown in Fig. \ref{Fig. 5}(b).

The Miller capacitors, $C\textsubscript{\textit{m}1}$ and $C\textsubscript{\textit{m}2}$, are employed in Figs. \ref{Fig. 5}(a) and (b). However, to simplify the circuit analysis, $C\textsubscript{\textit{m}1}$ and $C\textsubscript{\textit{m}2}$ are first neglected. Extending Eq. (1) to the three-stage feedforward structure while neglecting $C\textsubscript{\textit{d}}$, the overall transfer function of Fig. \ref{Fig. 5}(b) is given by

\begin{align}
	\nonumber
	\frac{V\textsubscript{\textit{out}}}{V\textsubscript{\textit{in}}}(s)\bigg|\textsubscript{3-\textit{stage}}&=\frac{A\textsubscript{2-\textit{stage}}A\textsubscript{\textit{v}3}}{\Big(1+\frac{s}{\omega^\prime\textsubscript{\textit{p}1}}\Big)\Big(1+\frac{s}{\omega^\prime\textsubscript{\textit{p}2}}\Big)}\\
	&\hspace{10pt}+\frac{A\textsubscript{\textit{vF}2}}{\Big(1+\frac{s}{\omega^\prime\textsubscript{\textit{p}2}}\Big)}
\end{align}

\noindent
where $A\textsubscript{2-\textit{stage}}$ is an approximated voltage gain of the internal two-stage structure, $A\textsubscript{\textit{v}3}$ is the output stage gain of the main path, $A\textsubscript{\textit{vF}2}$ is the output stage gain of the feedforward path, $\omega^\prime\textsubscript{\textit{p}1}$ is a pole at the output of the internal two-stage structure, and $\omega^\prime\textsubscript{\textit{p}2}$ is an output pole. Neglecting $C\textsubscript{\textit{d}}$ for simplicity, each stage has a gain as follows: $A\textsubscript{2-\textit{stage}}=g\textsubscript{\textit{m}1}R\textsubscript{\textit{o}1}g\textsubscript{\textit{m}2}(R\textsubscript{\textit{o}2}||R\textsubscript{\textit{oF}1})$, $A\textsubscript{\textit{v}3}=g\textsubscript{\textit{m}3}(R\textsubscript{\textit{o}3}||R\textsubscript{\textit{oF}2})$, and $A\textsubscript{\textit{vF}2}=g\textsubscript{\textit{mF}2}(R\textsubscript{\textit{o}3}||R\textsubscript{\textit{oF}2})$. To obtain poles and zero in Fig. \ref{Fig. 5}(b), Eq. (3) can be rewritten as

\begin{align}
	\frac{\big(A\textsubscript{2-\textit{stage}}A\textsubscript{\textit{v}3}+A\textsubscript{\textit{vF}2}\big)\bigg[1+\frac{A\textsubscript{\textit{vF}2}s}{\big(A\textsubscript{2-\textit{stage}}A\textsubscript{\textit{v}3}+A\textsubscript{\textit{vF}2}\big)\omega^\prime\textsubscript{\textit{p}1}}\bigg]}{\Big(1+\frac{s}{\omega^\prime\textsubscript{\textit{p}1}}\Big)\Big(1+\frac{s}{\omega^\prime\textsubscript{\textit{p}2}}\Big)}
\end{align}

In Eq. (4), the gain at low frequencies is $g\textsubscript{\textit{m}1}R\textsubscript{\textit{o}1}g\textsubscript{\textit{m}2}(R\textsubscript{\textit{o}2}||R\textsubscript{\textit{oF}1})g\textsubscript{\textit{m}3}(R\textsubscript{\textit{o}3}||R\textsubscript{\textit{oF}2})+g\textsubscript{\textit{mF}2}(R\textsubscript{\textit{o}3}||R\textsubscript{\textit{oF}2})$. This gain is approximated by $g\textsubscript{\textit{m}1}R\textsubscript{\textit{o}1}g\textsubscript{\textit{m}2}(R\textsubscript{\textit{o}2}||R\textsubscript{\textit{oF}1})g\textsubscript{\textit{m}3}(R\textsubscript{\textit{o}3}||R\textsubscript{\textit{oF}2})$. Two poles are obtained as $\omega^\prime\textsubscript{\textit{p}1}=1/(R\textsubscript{\textit{o}2}||R\textsubscript{\textit{oF}1})(C\textsubscript{\textit{o}2}+C\textsubscript{\textit{oF}1})$ and $\omega^\prime\textsubscript{\textit{p}2}=1/(R\textsubscript{\textit{o}3}||R\textsubscript{\textit{oF}2})(C\textsubscript{\textit{o}3}+C\textsubscript{\textit{oF}2})$. The zero by the feedforward path of $g\textsubscript{\textit{mF}2}$ is expressed as

\begin{align}
	\nonumber
	\omega^\prime\textsubscript{\textit{z}1}&=\omega^\prime\textsubscript{\textit{p}1}\bigg(\frac{A\textsubscript{2-\textit{stage}}A\textsubscript{\textit{v}3}}{A\textsubscript{\textit{vF}2}}+1\bigg)\\
	&=\omega^\prime\textsubscript{\textit{p}1}\bigg[\frac{g\textsubscript{\textit{m}1}R\textsubscript{\textit{o}1}g\textsubscript{\textit{m}2}(R\textsubscript{\textit{o}2}||R\textsubscript{\textit{oF}1})g\textsubscript{\textit{m}3}}{g\textsubscript{\textit{mF}2}}+1\bigg]
\end{align}

\begin{figure}[h!]
	\centering
	\includegraphics[scale=0.5]{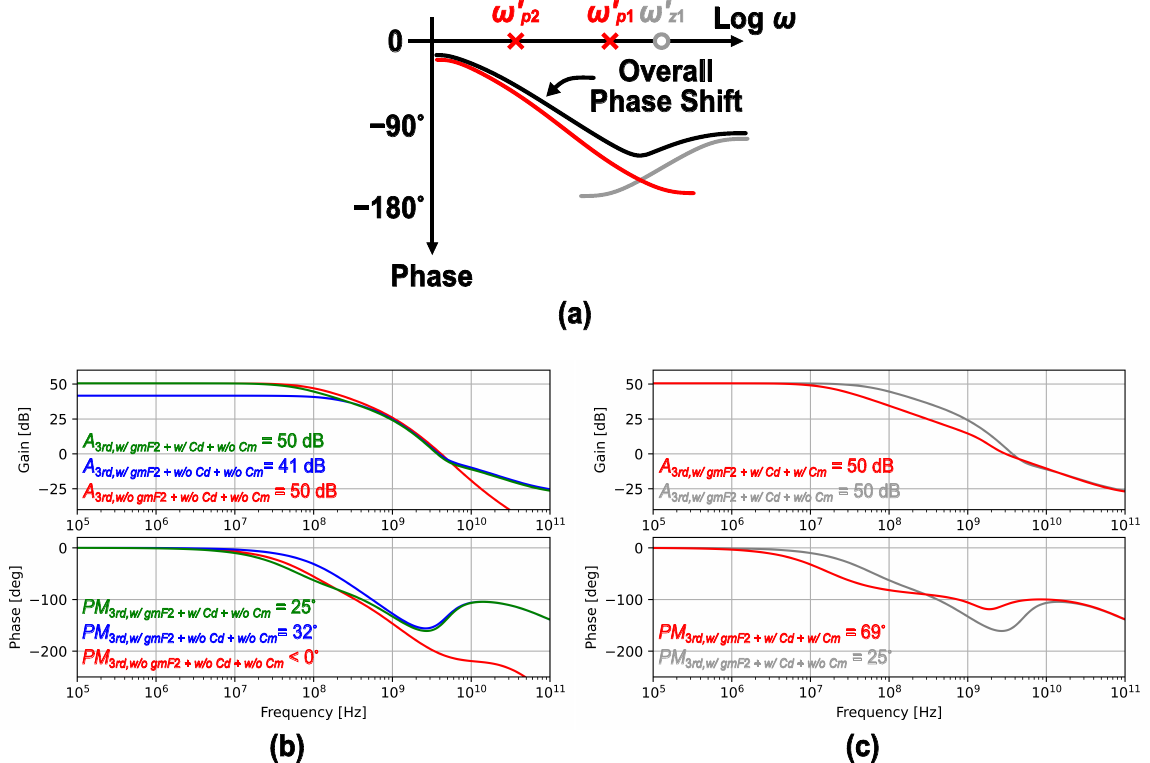}
	\caption{(a) Phase shift of the three-stage feedforward OTA, (b) Bode plots considering feedforward path and $C\textsubscript{\textit{d}}$, and (c) Bode plots considering $C\textsubscript{\textit{m}}$.}
	\label{Fig. 6}
\end{figure}

In Fig. \ref{Fig. 5}(b), considering $C\textsubscript{\textit{d}}$ at low frequencies, the output impedance of the feedforward path becomes high. At low frequencies, the output impedance is therefore dominated by $R\textsubscript{\textit{o}3}$ and $C\textsubscript{\textit{o}3}$ while ignoring $R\textsubscript{\textit{oF}2}$ and $C\textsubscript{\textit{oF}2}$. Accordingly, the voltage gain from Eq. (4) can be expressed as $g\textsubscript{\textit{m}1}R\textsubscript{\textit{o}1}g\textsubscript{\textit{m}2}(R\textsubscript{\textit{o}2}||R\textsubscript{\textit{oF}1})g\textsubscript{\textit{m}3}R\textsubscript{\textit{o}3}$. Also, the dominant pole $\omega^\prime\textsubscript{\textit{p}2}$, which is in a relatively lower frequency range than $\omega^\prime\textsubscript{\textit{p}1}$ and $\omega^\prime\textsubscript{\textit{z}1}$, changes from $1/(R\textsubscript{\textit{o}3}||R\textsubscript{\textit{oF}2})(C\textsubscript{\textit{o}3}+C\textsubscript{\textit{oF}2})$ to $1/(R\textsubscript{\textit{o}3}C\textsubscript{\textit{o}3})$. As the frequency increases, $C\textsubscript{\textit{d}}$ becomes almost a short circuit. Therefore, $\omega^\prime\textsubscript{\textit{p}1}$ and $\omega^\prime\textsubscript{\textit{z}1}$, which are in the relatively higher frequency range, remain similar to the state in the absence of $C\textsubscript{\textit{d}}$.

Recall that $\omega\textsubscript{\textit{z}1}=\omega\textsubscript{\textit{p}1}[(g\textsubscript{\textit{m}1}R\textsubscript{\textit{o}1}g\textsubscript{\textit{m}2}/g\textsubscript{\textit{mF}1})+1]$ from the two-stage feedforward OTA. If $g\textsubscript{\textit{mF}1}$ is large enough, $\omega\textsubscript{\textit{z}1}$ can be designed as $\omega\textsubscript{\textit{p}1}$, and $\omega\textsubscript{\textit{p}1}$ can be canceled out by $\omega\textsubscript{\textit{z}1}$. Similar to this operation, in Eq. (5), $\omega^\prime\textsubscript{\textit{z}1}$ can be designed as $\omega^\prime\textsubscript{\textit{p}1}$ if $g\textsubscript{\textit{mF}2}$ is large enough. However, in practical design, there is a limit to increasing $g\textsubscript{\textit{mF}2}$ and $g\textsubscript{\textit{m}1}R\textsubscript{\textit{o}1}g\textsubscript{\textit{m}2}(R\textsubscript{\textit{o}2}||R\textsubscript{\textit{oF}1})$ is also large, so $\omega^\prime\textsubscript{\textit{z}1}$ is generally designed to be larger than $\omega^\prime\textsubscript{\textit{p}1}$. Fig. \ref{Fig. 6}(a) shows the phase shift of the three-stage feedforward OTA and the phase peaking is generated due to the difference between $\omega^\prime\textsubscript{\textit{p}1}$ and $\omega^\prime\textsubscript{\textit{z}1}$.

Fig. \ref{Fig. 6}(b) presents the simulated Bode plots of the three-stage feedforward OTA according to $g\textsubscript{\textit{mF}2}$ and $C\textsubscript{\textit{d}}$. By including the feedforward amplifier $g\textsubscript{\textit{mF}2}$ (blue line), the PM is improved but the difference between $\omega^\prime\textsubscript{\textit{p}1}$ and $\omega^\prime\textsubscript{\textit{z}1}$ accompanies the phase peaking. Also, employing $C\textsubscript{\textit{d}}$ improves the voltage gain at low frequencies (green line). To further improve the PM, the Miller capacitors, $C\textsubscript{\textit{m}1}$ and $C\textsubscript{\textit{m}2}$ shown in Fig. \ref{Fig. 5}, are employed. As shown in Fig. \ref{Fig. 6}(c), the PM is improved through Miller compensation \cite{r4}, which exchanges the positions of $\omega^\prime\textsubscript{\textit{p}1}$ and $\omega^\prime\textsubscript{\textit{p}2}$, but the bandwidth is sacrificed. $\omega^\prime\textsubscript{\textit{z}1}$ also moves slightly toward the origin as $\omega^\prime\textsubscript{\textit{p}1}$ approaches the origin. In this three-stage OTA design, $C\textsubscript{\textit{d}}$ is set to 2 pF, and $C\textsubscript{\textit{m}1}$ and $C\textsubscript{\textit{m}2}$ are set to 50 fF. This means that $C\textsubscript{\textit{d}}$ makes the low-impedance path than $C\textsubscript{\textit{m}1}$ and $C\textsubscript{\textit{m}2}$. Therefore, the OTA is still dominated by the feedforward path than the Miller compensation path at high frequencies, which means that Eq. (5) is still valid to describe $\omega^\prime\textsubscript{\textit{z}1}$. Comparing the three-stage feedforward OTA and the two-stage feedforward OTA, the voltage gain improves as the number of stages increases. However, the three-stage structure shows a rapid phase change because $\omega^\prime\textsubscript{\textit{z}1}$ is not completely designed as $\omega^\prime\textsubscript{\textit{p}1}$ in an actual circuit design.

\begin{figure}[h!]
	\centering
	\includegraphics[scale=0.5]{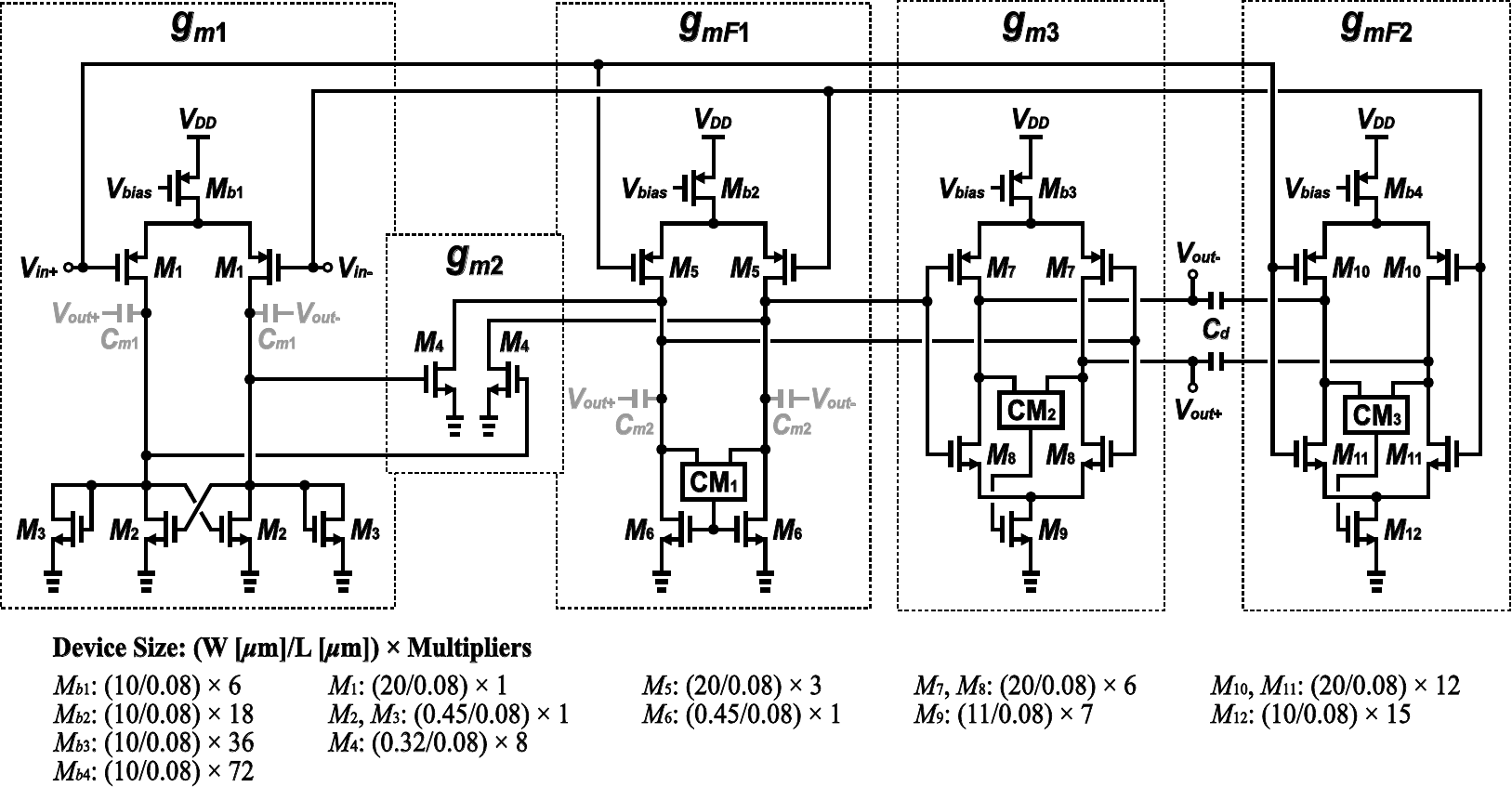}
	\caption{Fully differential three-stage feedforward OTA.\hspace{50pt}}
	\label{Fig. 7}
\end{figure}

Figs. \ref{Fig. 6}(b) and (c) are obtained using the fully differential three-stage feedforward OTA shown in Fig. \ref{Fig. 7}. The bias block and common-mode feedback are the same as those shown in Figs. \ref{Fig. 3}(b) and (c). The output load impedance is modeled as 10 M$\Omega||$2 pF on $V\textsubscript{\textit{out}--}$ and $V\textsubscript{\textit{out}+}$. The supply voltage $V\textsubscript{\textit{DD}}$ is set to 1.2 V. The common-mode voltages of $V\textsubscript{\textit{out}+/--}$ and $V\textsubscript{\textit{in}+/--}$ are set to 0.6 V. The current consumption of each stage is as follows: $g\textsubscript{\textit{m}1}$ (274 $\mu$A), $g\textsubscript{\textit{m}2}$ (794 $\mu$A), $g\textsubscript{\textit{mF}1}$ (821 $\mu$A), $g\textsubscript{\textit{m}3}$ (1.64 mA), $g\textsubscript{\textit{mF}2}$ (3.28 mA), and $CM\textsubscript{\textit{n}=1,2,3}$ (104 $\mu$A). Note that each $CM\textsubscript{\textit{n}=1,2,3}$ sets the common-mode voltage using $V\textsubscript{\textit{cm}}$ (= 0.6 V) as shown in Fig. \ref{Fig. 3}(c).

\singlespacing
\singlespacing

\subsection{Analysis of Four-Stage Feedforward Structure}
Recall that $\omega\textsubscript{\textit{z}1}=\omega\textsubscript{\textit{p}1}\big[(g\textsubscript{\textit{m}1}R\textsubscript{\textit{o}1}g\textsubscript{\textit{m}2}/g\textsubscript{\textit{mF}1})+1\big]$ in the two-stage OTA, then the two-stage feedforward OTA can be approximated as a one-pole system if $\omega\textsubscript{\textit{z}1}=\omega\textsubscript{\textit{p}1}$ by making $g\textsubscript{\textit{mF}1}$ large enough. By employing the approximated model of the two-stage feedforward OTA, the three-stage design also can be approximated as shown in Fig. \ref{Fig. 5}(b).

Then, the zero of $\omega^\prime\textsubscript{\textit{z}1}$ is obtained as $\omega^\prime\textsubscript{\textit{p}1}\big[(g\textsubscript{\textit{m}1}R\textsubscript{\textit{o}1}g\textsubscript{\textit{m}2}(R\textsubscript{\textit{o}2}||R\textsubscript{\textit{oF}1})g\textsubscript{\textit{m}3}/g\textsubscript{\textit{mF}2})+1\big]$ as shown in Eq. (5). Similar to $\omega\textsubscript{\textit{z}1}$ of the two-stage design, if $g\textsubscript{\textit{mF}2}$ is large enough, $\omega^\prime\textsubscript{\textit{z}1}$ becomes $\omega^\prime\textsubscript{\textit{p}1}$, and $\omega^\prime\textsubscript{\textit{p}1}$ can be canceled out by $\omega^\prime\textsubscript{\textit{z}1}$. Therefore, theoretically, the three-stage OTA can be simplified into the two-pole system by making $g\textsubscript{\textit{mF}1}$ large and further simplified into the one-pole system by making $g\textsubscript{\textit{mF}2}$ sufficiently large, which is described through SOPA as shown in Fig. \ref{Fig. 4}(b).

\begin{figure}[h!]
	\centering
	\includegraphics[scale=0.5]{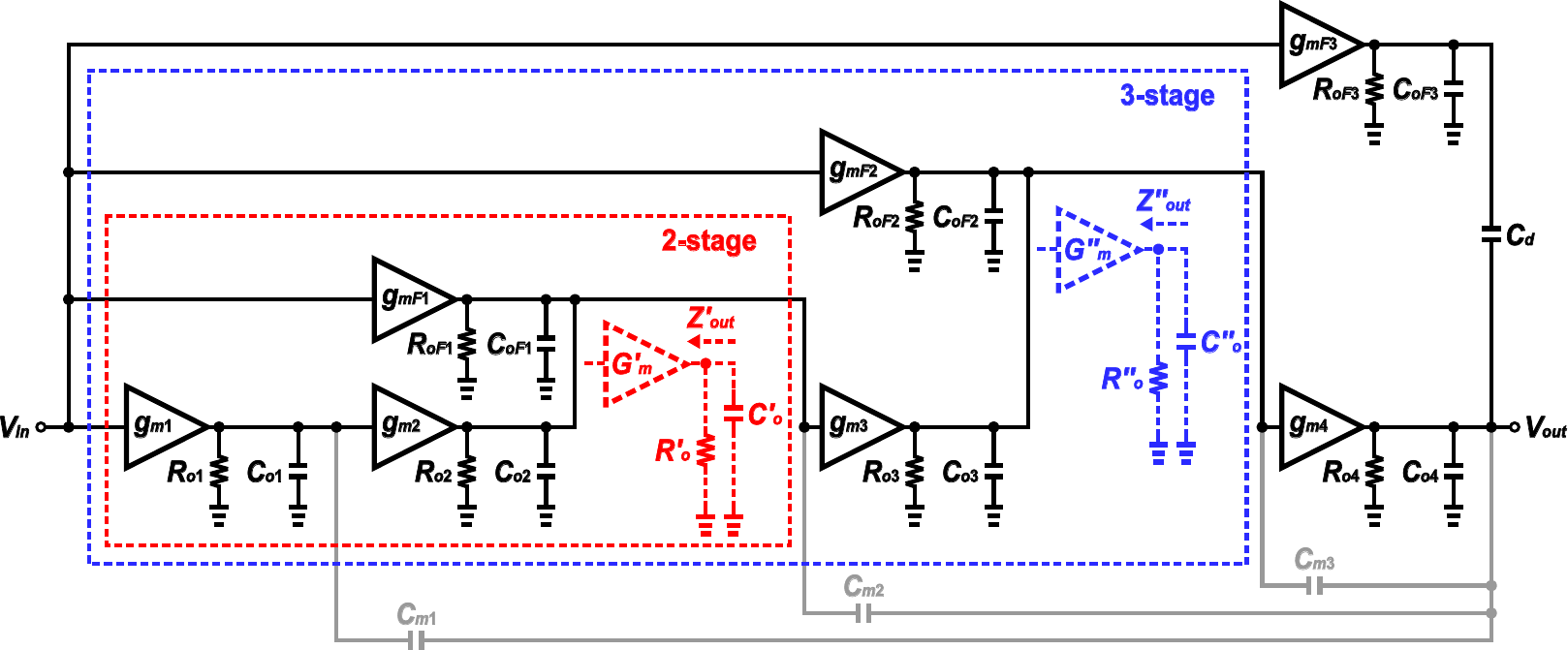}
	\caption{Four-stage feedforward OTA with SOPA.}
	\label{Fig. 8}
\end{figure}

Fig. \ref{Fig. 8} shows the four-stage feedforward OTA including the approximated circuit model of the preceding stages through SOPA. Note that the Miller capacitors $C\textsubscript{\textit{m}1,2,3}$ are employed in the four-stage design, but will be first neglected in the analysis for simplicity and discussed in the Bode plot results. In the red box of Fig. \ref{Fig. 8}, if $g\textsubscript{\textit{mF}1}$ is large enough, the two-stage design can be simplified as the one-pole system employing $G^\prime\textsubscript{\textit{m}}=(g\textsubscript{\textit{m}1}R\textsubscript{\textit{o}1})g\textsubscript{\textit{m}2}$, $R^\prime\textsubscript{\textit{o}}=R\textsubscript{\textit{o}2}||R\textsubscript{\textit{oF}1}$, and $C^\prime\textsubscript{\textit{o}}=C\textsubscript{\textit{o}2}+C\textsubscript{\textit{oF}1}$. In the blue box of Fig. \ref{Fig. 8}, if $g\textsubscript{\textit{mF}1}$ and $g\textsubscript{\textit{mF}2}$ are large enough, the three-stage design can be simplified into the two-pole system and further simplified into the one-pole system by employing $G^{\prime\prime}\textsubscript{\textit{m}}$, $R^{\prime\prime}\textsubscript{\textit{o}}$, and $C^{\prime\prime}\textsubscript{\textit{o}}$. In Eq. (4), $G^{\prime\prime}\textsubscript{\textit{m}}$ is approximately obtained as $g\textsubscript{\textit{m}1}R\textsubscript{\textit{o}1}g\textsubscript{\textit{m}2}(R\textsubscript{\textit{o}2}||R\textsubscript{\textit{oF}1})g\textsubscript{\textit{m}3}$. $R^{\prime\prime}\textsubscript{\textit{o}}$ and $C^{\prime\prime}\textsubscript{\textit{o}}$ are obtained as $R\textsubscript{\textit{o}3}||R\textsubscript{\textit{oF}2}$ and $C\textsubscript{\textit{o}3}+C\textsubscript{\textit{oF}2}$, respectively. 

Leveraging approximated one-pole models of the two- and three-stage OTAs while neglecting $C\textsubscript{\textit{d}}$ in Fig. \ref{Fig. 8}, the transfer function of the four-stage OTA can be given by

\begin{align}
	\nonumber
	\frac{V\textsubscript{\textit{out}}}{V\textsubscript{\textit{in}}}(s)\bigg|\textsubscript{4-\textit{stage}}&=\frac{A\textsubscript{3-\textit{stage}}A\textsubscript{\textit{v}4}}{\Big(1+\frac{s}{\omega^{\prime\prime}\textsubscript{\textit{p}1}}\Big)\Big(1+\frac{s}{\omega^{\prime\prime}\textsubscript{\textit{p}2}}\Big)}\\
	&\hspace{10pt}+\frac{A\textsubscript{\textit{vF}3}}{\Big(1+\frac{s}{\omega^{\prime\prime}\textsubscript{\textit{p}2}}\Big)}\\
	\nonumber
	\\
	\nonumber
	&=\big(A\textsubscript{3-\textit{stage}}A\textsubscript{\textit{v}4}+A\textsubscript{\textit{vF}3}\big)\\
	&\hspace{10pt}\times\frac{\bigg[1+\frac{A\textsubscript{\textit{vF}3}s}{\big(A\textsubscript{3-\textit{stage}}A\textsubscript{\textit{v}4}+A\textsubscript{\textit{vF}3}\big)\omega^{\prime\prime}\textsubscript{\textit{p}1}}\bigg]}{\Big(1+\frac{s}{\omega^{\prime\prime}\textsubscript{\textit{p}1}}\Big)\Big(1+\frac{s}{\omega^{\prime\prime}\textsubscript{\textit{p}2}}\Big)}
\end{align}

\noindent
where $A\textsubscript{3-\textit{stage}}=G^{\prime\prime}\textsubscript{\textit{m}}R^{\prime\prime}\textsubscript{\textit{o}}$, $A\textsubscript{\textit{v}4}=g\textsubscript{\textit{m}4}(R\textsubscript{\textit{o}4}||R\textsubscript{\textit{oF}3})$, and $A\textsubscript{\textit{vF}3}=g\textsubscript{\textit{mF}3}(R\textsubscript{\textit{o}4}||R\textsubscript{\textit{oF}3})$. The poles of the three- and four-stage outputs are expressed as $\omega^{\prime\prime}\textsubscript{\textit{p}1}=1/(R\textsubscript{\textit{o}3}||R\textsubscript{\textit{oF}2})(C\textsubscript{\textit{o}3}+C\textsubscript{\textit{oF}2})$ and $\omega^{\prime\prime}\textsubscript{\textit{p}2}=1/(R\textsubscript{\textit{o}4}||R\textsubscript{\textit{oF}3})(C\textsubscript{\textit{o}4}+C\textsubscript{\textit{oF}3})$, respectively. In Eq. (7) obtained using SOPA, the zero of the four-stage feedforward OTA is given by

\begin{align}
	\nonumber
	\omega^{\prime\prime}\textsubscript{\textit{z}1}&=\omega^{\prime\prime}\textsubscript{\textit{p}1}\bigg(\frac{A\textsubscript{3-\textit{stage}}A\textsubscript{\textit{v}4}}{A\textsubscript{\textit{vF}3}}+1\bigg)\\
	\nonumber
	&=\omega^{\prime\prime}\textsubscript{\textit{p}1}\bigg[\cdot\bigg]\\
	\nonumber
	&\hspace{35pt}\Downarrow\\
	&\bigg[\frac{g\textsubscript{\textit{m}1}R\textsubscript{\textit{o}1}g\textsubscript{\textit{m}2}(R\textsubscript{\textit{o}2}||R\textsubscript{\textit{oF}1})g\textsubscript{\textit{m}3}(R\textsubscript{\textit{o}3}||R\textsubscript{\textit{oF}2})g\textsubscript{\textit{m}4}}{g\textsubscript{\textit{mF}3}}+1\bigg]
\end{align}

Table 1 summarizes the zeros according to the number of stages. Recall that $\omega^\prime\textsubscript{\textit{z}1}$ and $\omega^{\prime\prime}\textsubscript{\textit{z}1}$ are obtained assuming that $g\textsubscript{\textit{mF}1}$ and $g\textsubscript{\textit{mF}2}$ are large enough. Similar to two- and three-stage OTAs, for improving the PM in the four-stage feedforward OTA, $g\textsubscript{\textit{mF}3}$ must also be large enough to cancel out $\omega^{\prime\prime}\textsubscript{\textit{p}1}$. From the equations of $\omega\textsubscript{\textit{z}1}$, $\omega^\prime\textsubscript{\textit{z}1}$, and $\omega^{\prime\prime}\textsubscript{\textit{z}1}$ as summarized in Table 1, it can be concluded that the transconductance of the feedforward path must be sufficiently large to secure the PM. Assuming that $g\textsubscript{\textit{mF}1}$ and $g\textsubscript{\textit{mF}2}$ are large enough, Eq. (7) can be rewritten as

\begin{align}
	\nonumber
	\frac{V\textsubscript{\textit{out}}}{V\textsubscript{\textit{in}}}(s)\bigg|\textsubscript{4-\textit{stage}}&=\frac{A\textsubscript{4-\textit{stage}}\Big(1+\frac{s}{\omega\textsubscript{\textit{z}1}}\Big)\Big(1+\frac{s}{\omega^\prime\textsubscript{\textit{z}1}}\Big)}{\Big(1+\frac{s}{\omega\textsubscript{\textit{p}1}}\Big)\Big(1+\frac{s}{\omega^\prime\textsubscript{\textit{p}1}}\Big)}\\
	&\hspace{10pt}\times\frac{\Big(1+\frac{s}{\omega^{\prime\prime}\textsubscript{\textit{z}1}}\Big)}{\Big(1+\frac{s}{\omega^{\prime\prime}\textsubscript{\textit{p}1}}\Big)\Big(1+\frac{s}{\omega^{\prime\prime}\textsubscript{\textit{p}2}}\Big)}
\end{align}

\noindent
where $A\textsubscript{4-\textit{stage}}$ can be approximated by $G^{\prime\prime}\textsubscript{\textit{m}}R^{\prime\prime}\textsubscript{\textit{o}}g\textsubscript{\textit{m}4}(R\textsubscript{\textit{o}4}||R\textsubscript{\textit{oF}3})$. The four-stage OTA shown in Fig. \ref{Fig. 8} consists of four poles and three zeros in Eq. (9). Ideally, by making $g\textsubscript{\textit{mF}1}$ large enough, $[1+(s/\omega\textsubscript{\textit{p}1})]$ in the denominator can be canceled out by $[1+(s/\omega\textsubscript{\textit{z}1})]$ in the numerator. Similarly, $[1+(s/\omega^\prime\textsubscript{\textit{p}1})]$ in the denominator can be canceled out by $[1+(s/\omega^\prime\textsubscript{\textit{z}1})]$ in the numerator as $g\textsubscript{\textit{mF}2}$ becomes large enough. Eventually, the four-stage feedforward OTA can be converted into the one-pole system assuming that $[1+(s/\omega^{\prime\prime}\textsubscript{\textit{p}1})]$ in the denominator is canceled out by $[1+(s/\omega^{\prime\prime}\textsubscript{\textit{z}1})]$ in the numerator as $g\textsubscript{\textit{mF}3}$ becomes sufficiently large. The approximation procedure of Eq. (9) described above is illustrated by SOPA in Fig. \ref{Fig. 4}(b). As mentioned above, to satisfy the condition that the poles are canceled out by the zeros, all feedforward stages need to be designed to have sufficiently large transconductances while sustaining $g\textsubscript{\textit{mF}1} \ll g\textsubscript{\textit{mF}2} \ll g\textsubscript{\textit{mF}3}$ as shown in Table 1. However, as the number of stages increases, the burden on the transconductance of the feedforward path increases for canceling out poles, which is associated with design area and power consumption.

\singlespacing
\begin{table}[h!]
		\centering
	\begin{tabular}{p{20mm} | p{110mm}}
		\multicolumn{2}{c}{\textbf{Table 1.} Zeros according to the number of stages}\\
		\hline
		\hline
		\textbf{\addstackgap[5pt]Order}&\textbf{Zero}\\
		\hline
		\addstackgap[8pt]{2-stage}&\hspace{10pt}$\omega\textsubscript{\textit{z}1}=\omega\textsubscript{\textit{p}1}\Big(\frac{g\textsubscript{\textit{m}1}R\textsubscript{\textit{o}1}g\textsubscript{\textit{m}2}}{g\textsubscript{\textit{mF}1}}+1\Big)$\\
		\hline
		\addstackgap[12pt]{3-stage}&\hspace{10pt}$\omega^\prime\textsubscript{\textit{z}1}=\omega^\prime\textsubscript{\textit{p}1}\bigg[\frac{g\textsubscript{\textit{m}1}R\textsubscript{\textit{o}1}g\textsubscript{\textit{m}2}(R\textsubscript{\textit{o}2}||R\textsubscript{\textit{oF}1})g\textsubscript{\textit{m}3}}{g\textsubscript{\textit{mF}2}}+1\bigg]$\\
		\hline
		\addstackgap[12pt]{4-stage}&\hspace{10pt}$\omega^{\prime\prime}\textsubscript{\textit{z}1}=\omega^{\prime\prime}\textsubscript{\textit{p}1}\bigg[ \cdot \bigg]$\\
		{}&\hspace{71pt}$\Downarrow$\\
		\addstackgap[12pt]{}&\hspace{10pt}$\bigg[\frac{g\textsubscript{\textit{m}1}R\textsubscript{\textit{o}1}g\textsubscript{\textit{m}2}(R\textsubscript{\textit{o}2}||R\textsubscript{\textit{oF}1})g\textsubscript{\textit{m}3}(R\textsubscript{\textit{o}3}||R\textsubscript{\textit{oF}2})g\textsubscript{\textit{m}4}}{g\textsubscript{\textit{mF}3}}+1\bigg]$\\
		\hline
		\hline
	\end{tabular}
	\label{Table 1}
\end{table}
\singlespacing
\singlespacing

In other words, to cancel out a pole by a zero in each stage, the transconductance ratio between the output stages of each order must be greater than the gain of the previous stage. Therefore, the following conditions must be satisfied:

\begin{align}
	(g\textsubscript{\textit{mF}1}/g\textsubscript{\textit{m}2}) &\gg g\textsubscript{\textit{m}1}R\textsubscript{\textit{o}1}\\
	(g\textsubscript{\textit{mF}2}/g\textsubscript{\textit{m}3}) &\gg g\textsubscript{\textit{m}1}R\textsubscript{\textit{o}1}g\textsubscript{\textit{m}2}(R\textsubscript{\textit{o}2}||R\textsubscript{\textit{oF}1})\\
	(g\textsubscript{\textit{mF}3}/g\textsubscript{\textit{m}4}) &\gg g\textsubscript{\textit{m}1}R\textsubscript{\textit{o}1}g\textsubscript{\textit{m}2}(R\textsubscript{\textit{o}2}||R\textsubscript{\textit{oF}1})g\textsubscript{\textit{m}3}(R\textsubscript{\textit{o}3}||R\textsubscript{\textit{oF}2})
\end{align}

\noindent
However, considering DC bias conditions, power consumption, and design area, it becomes more difficult to satisfy the above conditions as the number of stages increases. For these reasons, it becomes difficult to cancel out a pole that is generated as the number of stages increases, ultimately leading to the degradation of stability.

\begin{figure}[h!]
	\centering
	\includegraphics[scale=0.5]{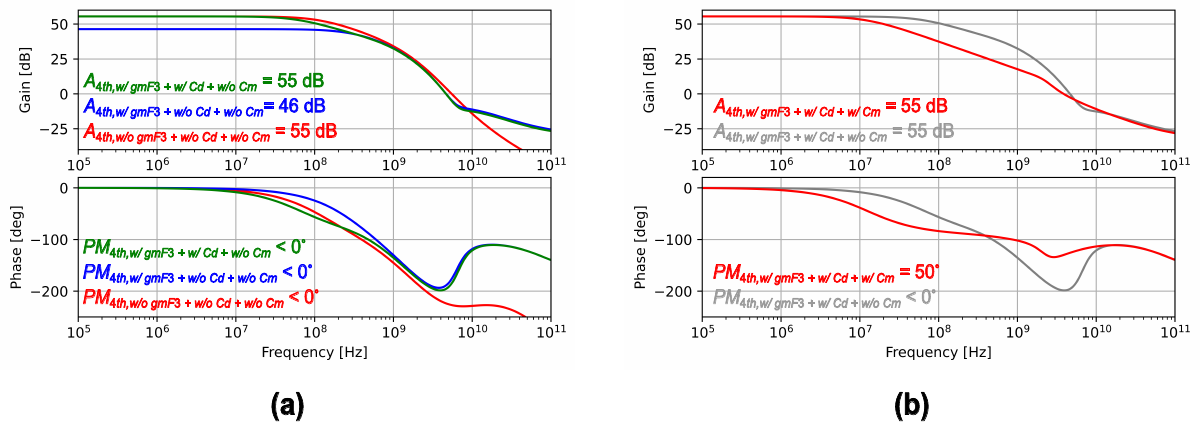}
	\caption{(a) Bode plots of the four-stage feedforward OTA without the Miller capacitors and (b) Bode plots employing the Miller capacitors.}
	\label{Fig. 9}
\end{figure}

Fig. \ref{Fig. 9}(a) shows the simulated Bode plots obtained using the four-stage feedforward OTA without Miller compensation. Employing the feedforward path with $g\textsubscript{\textit{mF}3}$ and without $C\textsubscript{\textit{d}}$ alleviates the phase shift at high frequencies compared to the design without the feedforward path (blue and red lines in Fig. \ref{Fig. 9}(a)). However, the zero is located at a higher frequency than the pole, and the zero does not completely cancel out the pole, leading to the PM degradation. As shown in the green line in Fig. \ref{Fig. 9}(a), by isolating the output feedforward path using $C\textsubscript{\textit{d}}$, the voltage gain is improved from $G^{\prime\prime}\textsubscript{\textit{m}}R^{\prime\prime}\textsubscript{\textit{o}}g\textsubscript{\textit{m}4}(R\textsubscript{\textit{o}4}||R\textsubscript{\textit{oF}3})$ to $G^{\prime\prime}\textsubscript{\textit{m}}R^{\prime\prime}\textsubscript{\textit{o}}g\textsubscript{\textit{m}4}R\textsubscript{\textit{o}4}$ at low frequencies. Also, the dominant pole $\omega^{\prime\prime}\textsubscript{\textit{p}2}$ changes from $1/(R\textsubscript{\textit{o}4}||R\textsubscript{\textit{oF}3})(C\textsubscript{\textit{o}4}+C\textsubscript{\textit{oF}3})$ to $1/(R\textsubscript{\textit{o}4}C\textsubscript{\textit{o}4})$ as $C\textsubscript{\textit{d}}$ is employed. The other poles and zeros located at high frequencies are almost unaffected by $C\textsubscript{\textit{d}}$ since $C\textsubscript{\textit{d}}$ becomes almost a short circuit.

As shown in Fig. \ref{Fig. 9}(b), the PM can be further improved by employing the Miller capacitors $C\textsubscript{\textit{m}1,2,3}$ shown in Fig. \ref{Fig. 8}. Similar to Miller compensation in the two-stage design, the poles and zeros can be located closely together through pole splitting \cite{r4}. Therefore, the zero can effectively cancel out the pole that shifts toward high frequencies by pole splitting, resulting in an improved PM. However, the pole that moves toward the origin through pole splitting reduces the bandwidth. In this analysis, $C\textsubscript{\textit{m}1}$ and $C\textsubscript{\textit{m}3}$ are set to 10 fF, and $C\textsubscript{\textit{m}2}$ is set to 200 fF. Also, the load impedance is set to 10 M$\Omega||$2 pF on $V\textsubscript{\textit{out}}$ for this analysis. Note that the output load impedance is set to 10 M$\Omega||$2 pF in analyzing all multistage feedforward OTAs throughout this work, and all output stages are designed using large device size to achieve a large transconductance. Accordingly, when analyzing the OTAs without Miller capacitors, it is assumed that $\omega\textsubscript{\textit{p}2}$, $\omega^\prime\textsubscript{\textit{p}2}$, and $\omega^{\prime\prime}\textsubscript{\textit{p}2}$ are the dominant poles. 

\begin{figure}[h!]
	\centering
	\includegraphics[scale=0.5]{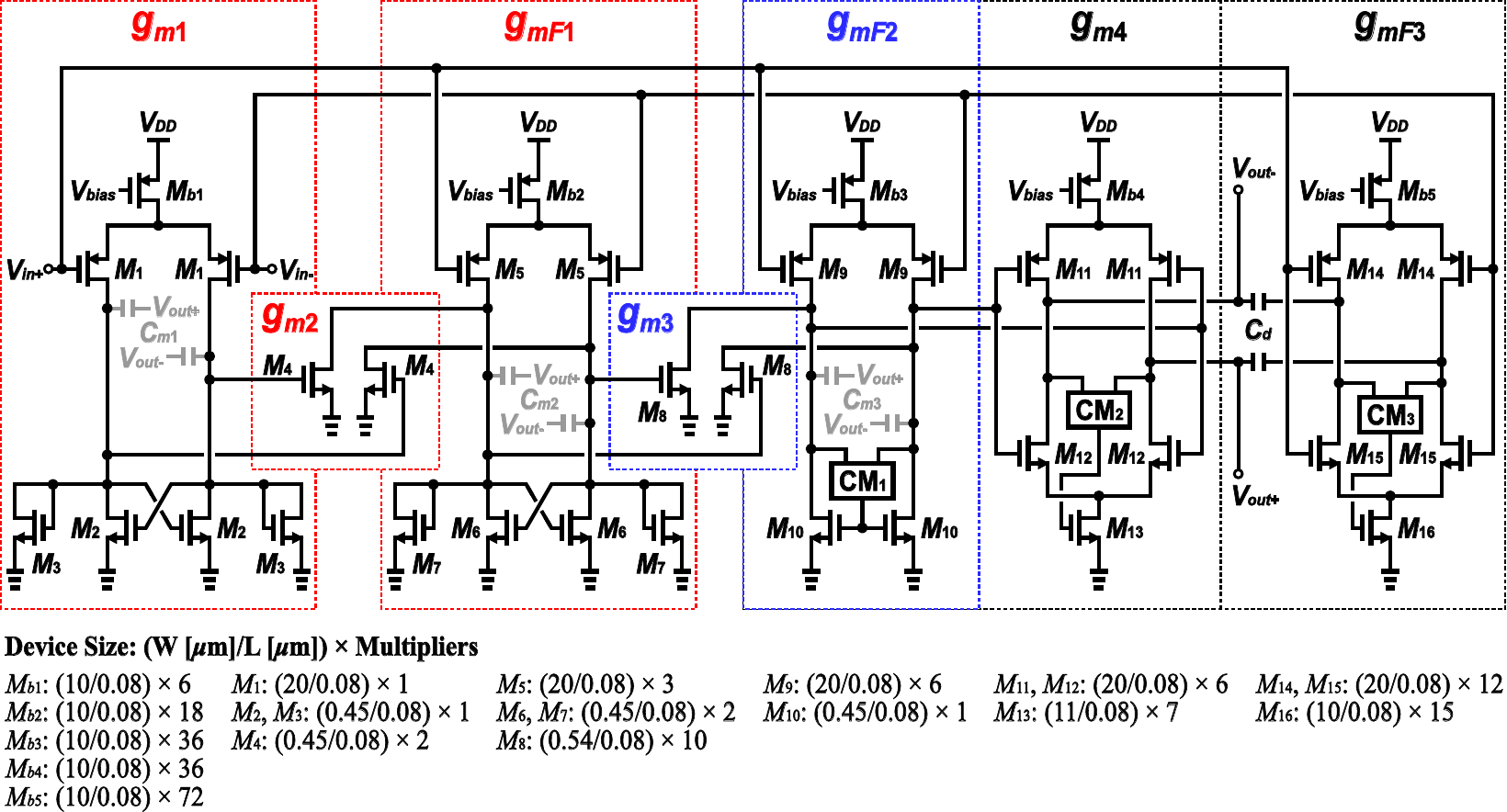}
	\caption{Fully differential four-stage feedforward OTA.\hspace{100pt}}
	\label{Fig. 10}
\end{figure}

Fig. \ref{Fig. 10} shows the fully differential feedforward OTA used to obtain the results shown in  Fig. \ref{Fig. 9}. In Fig. \ref{Fig. 10}, the bias block and common-mode feedbacks $CM\textsubscript{\textit{n=}1,2,3}$ are the same as the circuits used in two- and three-stage OTAs. The common-mode voltages of $V\textsubscript{\textit{in}+/--}$ and $V\textsubscript{\textit{out}+/--}$ are set to 0.6 V. $C\textsubscript{\textit{d}}$ is set to 2 pF. The current consumption of the four-stage feedforward OTA is as follows: $g\textsubscript{\textit{m}1}$ (273 $\mu$A), $g\textsubscript{\textit{m}2}$ (273 $\mu$A), $g\textsubscript{\textit{mF}1}$ (820 $\mu$A), $g\textsubscript{\textit{m}3}$ (1.61 mA), $g\textsubscript{\textit{mF}2}$ (1.64 mA), $g\textsubscript{\textit{m}4}$ (1.63 mA), $g\textsubscript{\textit{mF}3}$ (3.28 mA), and $CM\textsubscript{\textit{n=}1,2,3}$ (104 $\mu$A). The OTA is driven using a 1.2 V supply voltage.

Table 2 summarizes the overall transconductances used in this analysis according to the number of stages. Note that the transconductance ratio is defined as $g\textsubscript{\textit{mF}(k-1)}/g\textsubscript{\textit{m}(k)}\textsubscript{, k=2,3,4}$ in this analysis. As discussed above, $g\textsubscript{\textit{mF}(k-1)}/g\textsubscript{\textit{m}(k)}$ must be designed as large as possible to secure sufficient phase margin in a multistage feedforward OTA. However, as the OTA order increases, achieving sufficient $g\textsubscript{\textit{mF}(k-1)}/g\textsubscript{\textit{m}(k)}$ is a challenging task and becomes even more difficult when considering  DC bias conditions, power consumption, and design area. Therefore, the stability analysis is carried out by setting $g\textsubscript{\textit{mF}(k-1)}/g\textsubscript{\textit{m}(k)}$ to be larger than 2 in this work, as summarized in Table 2.

\singlespacing
\begin{table}[h!]
		\centering
	\begin{tabular}{p{15mm} | p{35mm} p{35mm} p{35mm}}
		\multicolumn{4}{c}{\textbf{Table 2.} Overall transconductances}\\
		\hline
		\hline
		\textbf{\addstackgap[5pt]Order}&{}&\textbf{Transconductance}&{}\\
		\hline
		\addstackgap[3pt]{2-stage}&$g\textsubscript{\textit{m}1}$&$g\textsubscript{\textit{m}2}$&$g\textsubscript{\textit{mF}1}$\\
		{}&(1.99 mS)&(22.12 mS)&(44.97 mS)\\
		\hline
		\addstackgap[3pt]{3-stage}&$g\textsubscript{\textit{m}1}$&$g\textsubscript{\textit{m}2}$&$g\textsubscript{\textit{mF}1}$\\
		{}&(1.99 mS)&(2.27 mS)&(5.97 mS)\\
		{}&{}&$g\textsubscript{\textit{m}3}$&$g\textsubscript{\textit{mF}2}$\\
		{}&{}&(28.04 mS)&(56.29 mS)\\
		\hline
		\addstackgap[3pt]{4-stage}&$g\textsubscript{\textit{m}1}$&$g\textsubscript{\textit{m}2}$&$g\textsubscript{\textit{mF}1}$\\
		{}&(1.99 mS)&(0.79 mS)&(5.96 mS)\\
		{}&{}&$g\textsubscript{\textit{m}3}$&$g\textsubscript{\textit{mF}2}$\\
		{}&{}&(4.76 mS)&(11.95 mS)\\
		{}&{}&$g\textsubscript{\textit{m}4}$&$g\textsubscript{\textit{mF}3}$\\
		{}&{}&(28.05 mS)&(56.28 mS)\\
		\hline
	\end{tabular}
	\label{Table 2}
\end{table}
\singlespacing
\singlespacing

\section{Conclusion}
In this work, the stability analysis of multistage feedforward OTAs is conducted by employing SOPA. In each multistage feedforward OTA, SOPA is conducted as follows: (1) the first two-stage feedforward structure is modeled as the one-pole system by assuming that the first feedforward transconductance is large enough, (2) the next stage is again approximated identically by leveraging the previous one-pole system and making the second feedforward transconductance sufficiently large, and (3) the approximation is successively conducted by employing the previous one-pole system and making the next feedforward transconductance large enough.

To ensure stability, each feedforward structure must satisfy Eq. (13).  

\begin{align}
	\nonumber
	\frac{g\textsubscript{\textit{mF}(k-1)}}{g\textsubscript{\textit{m}(k)}} &\gg g\textsubscript{\textit{m}1}R\textsubscript{\textit{o}1}\times\big[g\textsubscript{\textit{m}2}(R\textsubscript{\textit{o}2}||R\textsubscript{\textit{oF}1})\big] \times \cdot\cdot\cdot\\
	&\hspace{13pt}\cdot\cdot\cdot\times\big[g\textsubscript{\textit{m}(k-2)}(R\textsubscript{\textit{o}(k-2)}||R\textsubscript{\textit{oF}(k-3)})\big]\times\big[g\textsubscript{\textit{m}(k-1)}(R\textsubscript{\textit{o}(k-1)}||R\textsubscript{\textit{oF}(k-2)})\big]
\end{align} 

\noindent
where $k$ (= 2, 3, 4, ...) means the order of a multistage feedforward OTA, and $R\textsubscript{\textit{oF}(0)}$ is an infinite value. Eq. (13) is obtained by generalizing Eqs. (10), (11), and (12). Then, the generalized transfer function of a multistage feedforward OTA can be expressed as Eqs. (14) and (15).

\begin{align}
	\frac{V\textsubscript{\textit{out}}}{V\textsubscript{\textit{in}}}(s)\bigg|\textsubscript{(k)-\textit{stage}}&=\frac{A\textsubscript{(k)-\textit{stage}}\Bigg[\Big(1+\frac{s}{\omega\textsubscript{\textit{z}1}}\Big)\Big(1+\frac{s}{\omega^\prime\textsubscript{\textit{z}1}}\Big)\times\cdot\cdot\cdot\times\Big(1+\frac{s}{\omega^{\prime\prime\cdot\cdot\cdot\prime}\textsubscript{\textit{z}1}}\Big)\Bigg]}{\Bigg[\Big(1+\frac{s}{\omega\textsubscript{\textit{p}1}}\Big)\Big(1+\frac{s}{\omega^\prime\textsubscript{\textit{p}1}}\Big)\times\cdot\cdot\cdot\times\Big(1+\frac{s}{\omega^{\prime\prime\cdot\cdot\cdot\prime}\textsubscript{\textit{p}1}}\Big)\Bigg]\Big(1+\frac{s}{\omega^{\prime\prime\cdot\cdot\cdot\prime}\textsubscript{\textit{p}2}}\Big)}\\
	&=\frac{A\textsubscript{(k)-\textit{stage}}\Bigg[\Big(1+\frac{s}{\omega\textsubscript{\textit{z}1}}\Big)\times{\displaystyle\prod_{\#=\prime}^{\prime\prime\cdot\cdot\cdot\prime}\Big(1+\frac{s}{\omega^{\#}\textsubscript{\textit{z}1}}\Big)}\Bigg]}{\Bigg[\Big(1+\frac{s}{\omega\textsubscript{\textit{p}1}}\Big)\times{\displaystyle\prod_{\#=\prime}^{\prime\prime\cdot\cdot\cdot\prime}\Big(1+\frac{s}{\omega^{\#}\textsubscript{\textit{p}1}}\Big)}\Bigg]\Big(1+\frac{s}{\omega^{\prime\prime\cdot\cdot\cdot\prime}\textsubscript{\textit{p}2}}\Big)}
\end{align}

\noindent
where $A\textsubscript{(k)-\textit{stage}}$ means the voltage gain of the $k$-stage feedforward OTA at low frequencies. Eq. (14) is extended from the four-stage transfer function shown in Eq. (9) and further generalized as shown in Eq. (15). Also, Eqs. (14) and (15) describe the behavior of SOPA shown in Fig. \ref{Fig. 4}(b). Although designing a reliable OTA becomes difficult as the number of stages increases in actual circuit design, SOPA provides an intuitive way to analyze the stability of a multistage feedforward OTA.

\singlespacing
\singlespacing
\singlespacing
\singlespacing

\end{document}